%% file: paper.tex
\documentclass[lettersize,journal]{IEEEtran}
\usepackage{amsmath,amsfonts}
\usepackage{algorithmic}
\usepackage{algorithm}
\usepackage{array}
\usepackage[caption=false,font=normalsize,labelfont=sf,textfont=sf]{subfig}
\usepackage{textcomp}
\usepackage{stfloats}
\usepackage{url}
\usepackage{verbatim}
\usepackage{graphicx}
\usepackage{cite}
\usepackage{bm}
\usepackage{balance}
\usepackage{kotex}
\usepackage{xspace}
\usepackage{tikz}
\usepackage{enumitem}
\usepackage[hidelinks]{hyperref}
\usepackage{listings}
\newcommand{\niparagraph}[1]{\vspace{2pt}\noindent\textbf{#1}}

\newcommand{\simname}{LLMServingSim2.0\xspace}
\newcommand{\simnametitle}{LLMServingSim2.0\xspace}
\newcommand{\simnamesection}{LLMServingSim2.0\xspace}

\usepackage{booktabs}
\usepackage{colortbl}
\usepackage{xcolor}

\usepackage{booktabs}
\usepackage{amssymb} 
\usepackage{makecell}
\usepackage{threeparttable} 
\newcommand{\yescheck}{\ensuremath{\checkmark}}
\newcommand{\nocheck}{\ensuremath{\times}}
\newcommand{\semicheck}{\ensuremath{\triangle}}
\usepackage{titlesec}
\titlespacing*{\section} {0pt}{0.4ex}{0.4ex}
\titlespacing*{\subsection} {0pt}{0.4ex}{0.4ex}

\begin{document}

\title{\LARGE{\simnametitle: A Unified Simulator for Heterogeneous \\Hardware and Serving Techniques in LLM Infrastructure}}

\author{Jaehong Cho, Hyunmin Choi, \IEEEmembership{Member, IEEE}, and Jongse Park, \IEEEmembership{Senior Member, IEEE}

\vspace{-0.2cm}
\thanks{Received 4 October 2025; accepted 28 October 2025. This work was supported by Institute of Information \& communications Technology Planning \& Evaluation (IITP) (No.RS-2022-II221037, No.RS-2024-00396013), National Research Foundation of Korea (NRF) (No.RS-2024-00342148), Electronics and Telecommunications Research Institute (ETRI) (No.RS-2025-02305453), and funded by the Korea government (MSIT). This work was also partly supported by SK hynix. \textit{(Corresponding author: Jongse Park.)}}
\thanks{The authors are with the School of Computing, Korea Advanced Institute of Science and Technology, Daejeon 34141, South Korea (e-mail: \href{mailto:jhcho@casys.kaist.ac.kr}{jhcho@casys.kaist.ac.kr}; \href{mailto:hmchoi@casys.kaist.ac.kr}{hmchoi@casys.kaist.ac.kr}; \href{mailto:jspark@casys.kaist.ac.kr}{jspark@casys.kaist.ac.kr}).}
\thanks{Digital Object Identifier 10.1109/LCA.2025.3628325}
\vspace{-1cm}
}

\markboth{IEEE COMPUTER ARCHITECTURE LETTERS,~Vol.~24,~No.~2,~JULY-DECEMBER 2025}%
{Shell \MakeLowercase{\textit{et al.}}: A Sample Article Using IEEEtran.cls for IEEE Journals}

\IEEEpubid{
  \begin{minipage}{1\textwidth}
    \vspace{0.8cm}
    \copyright~2025 IEEE. Personal use of this material is permitted. 
    Permission from IEEE must be obtained for all other uses, in any current or future media,
    including reprinting/republishing this material for advertising or promotional purposes,
    creating new collective works, for resale or redistribution to servers or lists,
    or reuse of any copyrighted component of this work in other works.
  \end{minipage}
}

\maketitle

\input{body/abstract}
\input{body/intoduction}

\input{body/llmservingsim}

\input{body/evaluation}
\input{body/conclusion}

\bibliographystyle{IEEEtran}
\bibliography{paper}

\vfill

\end{document}

%% file: body/abstract.tex
\begin{abstract}

This paper introduces \simname, a system simulator designed for exploring heterogeneous hardware in large-scale LLM serving systems.
\simname addresses two key limitations of its predecessor: (1) integrating hardware models into system-level simulators is non-trivial due to the lack of a clear abstraction, and (2) existing simulators support only a narrow subset of serving techniques, leaving no infrastructure that captures the breadth of approaches in modern LLM serving.
To overcome these issues, \simname adopts trace-driven performance modeling, accompanied by an operator-level latency profiler, enabling the integration of new accelerators with a single command.
It further embeds up-to-date serving techniques while exposing flexible interfaces for request routing, cache management, and scheduling policies.
In a TPU case study, our profiler requires 18.5$\times$ fewer LoC and outperforms the predecessor's hardware-simulator integration, demonstrating \simname's low-effort hardware extensibility.
Our experiments further show that \simname reproduces GPU-based LLM serving with 1.9\% error, while maintaining practical simulation time, making it a comprehensive platform for both hardware developers and LLM service providers.

\end{abstract}
\vspace{-0.2cm}
\begin{IEEEkeywords}
Large language model (LLM), Inference serving, Heterogeneous hardware, Simulation infrastructure
\end{IEEEkeywords}


%% file: body/intoduction.tex
\section{Introduction}
\IEEEPARstart{W}{hile} large language model (LLM) serving technologies are advancing rapidly across both academia and industry, current efforts remain separated: (1) the systems community focuses on optimizing software for GPU-based scale-out infrastructures, while (2) the architecture community develops custom hardware accelerators without an easy way to evaluate them under these evolving software environments.
This separation leaves hardware developers with few options for validating their accelerators at deployment scale, while also making it difficult for LLM service providers to assess and adopt emerging hardware in their serving systems.
To address these challenges, LLMServingSim~\cite{llmservingsim} developed a system-level simulator for heterogeneous LLM serving, allowing researchers to plug in custom accelerators and explore software techniques.
However, there are two key limitations: (1) integrating new hardware models is non-trivial due to the lack of a clear abstraction, and (2) existing simulators~\cite{agrawal2024vidurlargescalesimulationframework,lin2025apexextensibledynamismawaresimulator,wu2025tokensimenablinghardwaresoftware} support only a narrow subset of serving techniques, leaving them insufficient to capture modern deployments.
Table~\ref{tab:sim-comparison} summarizes the differences among existing simulators and our work. 
Prior efforts target isolated aspects of LLM serving, but none provide a unified framework that combines parallelism, caching, offloading, and disaggregation with realistic memory and network modeling. 
As deployments increasingly employ these techniques together, a framework that captures both operator-level performance and system-level dynamics becomes essential.

\input{table/sim-compare}

To this end, we present \simname, a hardware/software co-simulation infrastructure that streamlines the integration of custom accelerators and incorporates modern serving techniques, allowing systematic evaluation of large-scale LLM serving.
Compared to its predecessor, \simname introduces four key advancements. 

\IEEEpubidadjcol

\begin{itemize}[leftmargin=1.0em]
    \item\textbf{Trace-driven performance modeling.} It supports trace-driven performance models, enabling researchers to easily perform system-level design explorations with diverse hardware and model configurations.    \item\textbf{Multi-instance and disaggregated serving.} It enables multi-instance simulation and P/D disaggregation, capturing complex interactions between heterogeneous instances.
    \item\textbf{MoE support with expert parallelism and offloading.} It provides MoE model support with expert parallelism and offloading, faithfully modeling network congestion, memory bandwidth contention, and expert routing dynamics.     
    \item\textbf{Memory-aware serving with prefix caching.} It introduces the first network and memory-aware simulation of prefix caching, including a customizable management policy.
\end{itemize} 

Unlike prior simulators, \simname brings these four capabilities together in a single flexible framework, enabling researchers to explore heterogeneous hardware, diverse serving topologies, and customizable policies. 
This unification bridges the gap between accuracy, scalability, and usability, establishing \simname as a comprehensive platform for large-scale LLM serving. 
\simname is available at {\url{https://github.com/casys-kaist/llmservingsim}}.

%% file: table/sim-compare.tex
\begin{table}[t]
\centering
\caption{Comparison of LLM serving simulators}
\label{tab:sim-comparison}
\resizebox{\linewidth}{!}{
\begin{tabular}{l|cc|ccc|ccc}
\toprule
\textbf{Simulator} 
& \multicolumn{2}{c|}{\textbf{Dissagg.}} 
& \multicolumn{3}{c|}{\textbf{Parallelism}} 
& \multicolumn{3}{c}{\textbf{Memory Model}} \\ 
\cmidrule(lr){2-3} \cmidrule(lr){4-6} \cmidrule(lr){7-9}
 & PD & AF & PP/TP & DP & EP & PA & PC & EO \\ 
\midrule
LLMServingSim~\cite{llmservingsim} & \nocheck & \yescheck & \yescheck & \nocheck & \nocheck & \yescheck & \nocheck & \nocheck \\
Vidur~\cite{agrawal2024vidurlargescalesimulationframework} & \nocheck & \nocheck & \yescheck & \yescheck & \nocheck & \semicheck & \nocheck & \nocheck \\
APEX~\cite{lin2025apexextensibledynamismawaresimulator} & \nocheck & \nocheck & \yescheck & \yescheck & \yescheck & \nocheck & \nocheck & \nocheck \\
TokenSim~\cite{wu2025tokensimenablinghardwaresoftware} & \yescheck & \nocheck & \yescheck & \semicheck & \nocheck & \yescheck & \semicheck & \nocheck \\
\textbf{Ours} & \yescheck & \yescheck & \yescheck & \yescheck & \yescheck & \yescheck & \yescheck & \yescheck \\ 
\bottomrule
\end{tabular}
}
\vspace{0.1em}

{\footnotesize
\raggedright
PD: Prefill/Decode Disaggregation, 
AF: Attention/FFN Disaggregation, 
PP/TP: Pipeline/Tensor Parallelism, 
DP: Data Parallelism, 
EP: Expert Parallelism, 
PA: PagedAttention, 
PC: Prefix Caching, 
EO: Expert Offloading.

\raggedright
\yescheck: fully supported,
\nocheck: not supported,
\semicheck: limited or partial support. \par
}
\vspace{-0.1em}

\end{table}

%% file: body/llmservingsim.tex
\section{\simnamesection}

\begin{figure*}[t]
  \centering
  \includegraphics[width=0.97\linewidth]{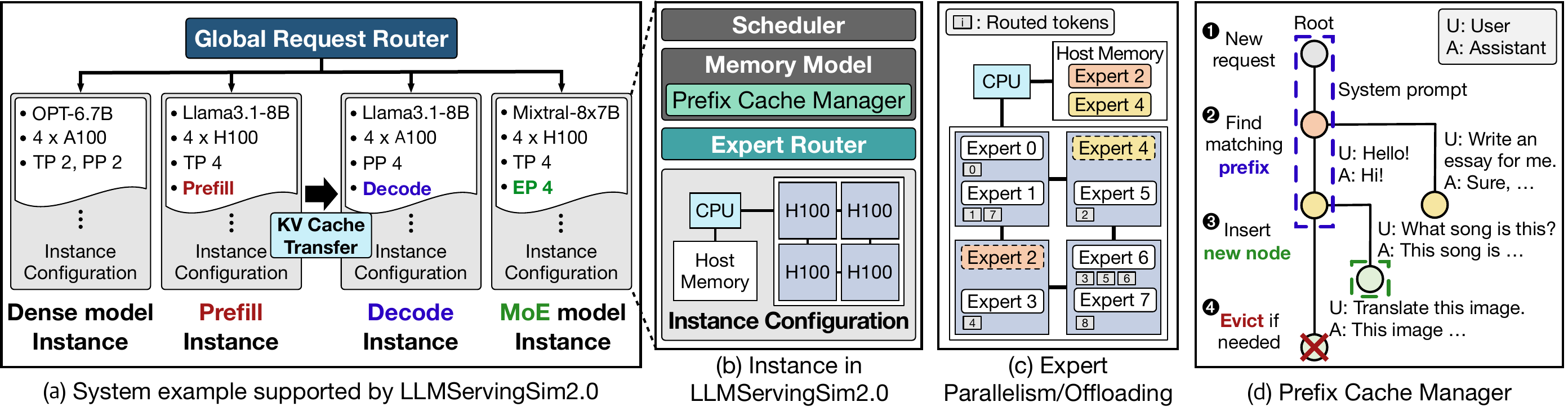}
  \vspace{-0.3em}
  \caption{Overview of LLMServingSim2.0 simulator.}
  \label{fig:overview}
  \vspace{-1.8em}
\end{figure*}

We design \simname, a next-generation system-level simulator for LLM inference that models a large-scale distributed LLM serving.
%
Fig.~\ref{fig:overview}(a) illustrates a simple example system supported by our simulator. 
\simname models heterogeneous instances along two dimensions: (1) model type (dense and MoE) and (2) serving structure (non-disaggregated and P/D disaggregated).
Each instance can be configured with distinct compute/memory resources, parallelism schemes, and network topologies, demonstrating the flexibility of our simulator.
%
%
%
%
%
%
%
Building on this flexibility, the \simname further supports advanced deployment scenarios, including multi-instance serving, P/D disaggregation, MoE expert parallelism/offloading (Fig.~\ref{fig:overview}(c)), and prefix caching (Fig.~\ref{fig:overview}(d)) within a unified simulation environment.


\subsection{Trace-driven Flexible Performance Modeling}
The key advantage of the original LLMServingSim~\cite{llmservingsim} was its ability to compose large-scale heterogeneous serving systems by integrating different hardware simulators, such as NPU and PIM.
However, porting a new hardware simulator to the platform can be a time-consuming and challenging task.
Moreover, due to the massive amount of repetitive computation and the autoregressive nature of LLMs, cycle-accurate hardware simulation remained slow even with computation reuse.

\niparagraph{Operator-level profiler.} To address these challenges, \simname introduces trace-driven performance modeling, enabling users to build performance studies directly from traces without extensive hardware simulator integration.
We developed \textit{operator-level profiler}, a PyTorch-based profiling tool that inserts hooks between LLM layers to measure layer-wise latency.
This profiler collects all the information required by \simname and, through validation against real execution, helps maintain high simulation accuracy.
With the profiler, users can analyze any model on their own hardware with a single-line command.
%
Compared to repeatedly simulating hardware in cycle-accurate mode, this approach is far more convenient and achieves 232$\times$ faster execution on average, while enabling flexible integration of diverse hardware platforms and models.
%

\subsection{Multi-instance and Disaggregation for Realistic Serving}

In large-scale LLM serving, hundreds of instances are deployed across rack-scale clusters or even entire data centers.
%
Such multi-instance execution is indispensable for emerging LLM-based applications 
%
and it introduces critical system-level effects, such as resource sharing, network contention, and global routing dynamics.
%
However, the original LLMServingSim only supported single-instance simulation, limiting its ability to capture the realities of modern large-scale serving.

\niparagraph{Global request router.} 
To tackle this challenge, \simname introduces a \textit{global request router} for multi-instance serving.
As illustrated in Fig.~\ref{fig:overview}(a), the global request router resides outside individual instances and manages all instances, dynamically dispatching requests according to the current resource status. 
Such routing can adapt to diverse factors, including load balancing, workload characteristics, and the state of prefix caches. 
Because routing policies critically affect overall performance, \simname provides customizable routing interfaces, enabling researchers to easily implement and explore new policies for large-scale LLM serving.

\niparagraph{Heterogeneous multi-instance.}
As shown in Fig.~\ref{fig:overview}(b), \simname realizes multi-instance support by replicating components that previously existed as global entities, thereby giving each instance its own scheduler and memory model. 
%
%
In real LLM services, multiple models often coexist, each with different resource demands, while server hardware can also vary. 
To capture this heterogeneity, \simname allows users to construct heterogeneous multi-instance systems. As depicted in Fig.~\ref{fig:overview}(a), each instance can be configured with different models, hardware types and counts, memory sizes and bandwidths, parallelism schemes, as well as distinct network topology. 
This flexibility enables \simname to model realistic serving infrastructures and allows researchers to explore a wider range of system designs.

\niparagraph{P/D disaggregation.}
Beyond hosting multiple heterogeneous instances, real-world deployments increasingly adopt P/D disaggregation~\cite{splitwise, distserve}, where the compute-intensive prefill stage and the memory-intensive decode stage are placed on separate clusters.
To support P/D disaggregation, \simname introduces dedicated prefill and decode instance types as shown in Fig.~\ref{fig:overview}(a). 
Once the prefill stage completes in a prefill instance, the request and its KV cache are forwarded to a decode instance for continuation.
%
%
%
Moreover, \simname exposes this KV cache transfer policy as a configurable option, allowing researchers to experiment with alternative strategies.
Building on phase-aware optimizations in various P/D-disaggregated serving systems \cite{splitwise, distserve}, LLMServingSim 2.0 supports systematic exploration of these and broader design choices across diverse system configurations.


\subsection{MoE Support: Expert Parallelism and Offloading}
LLM advances increasingly rely on compute and memory efficient designs to improve performance. Among various approaches, MoE architecture has gained significant traction as a high-performing, promising approach.
However, the original LLMServingSim only supports dense models and thus can not capture the emerging trend of MoE-based LLM serving.

\niparagraph{Expert router.}
To address this limitation, \simname introduces support for MoE models through expert routing and parallelism.
%
%
In MoE, a gate function routes tokens to the most suitable experts, thereby scaling up model capacity for higher accuracy while activating only a subset of experts per token to keep the computational cost nearly constant.
%
%
We design an \textit{expert router} (Fig.~\ref{fig:overview}(b)) that mimics the behavior of real gate functions. 
This router can be flexibly customized, enabling researchers to experiment with alternative routing strategies.

\niparagraph{Expert parallelism.}
Once tokens are routed to their designated experts, the next step is to distribute the experts across compute units, referred to as expert parallelism.
%
%
As illustrated in Fig.~\ref{fig:overview}(c), it partitions experts across compute units and routes tokens to their assigned experts.
In \simname, MoE instances are designed such that non-expert layers operate under conventional tensor or pipeline parallelism, while expert layers employ the expert router to dispatch tokens according to the assigned experts. 
Between these two types of layers, \simname models an all-to-all communication operation in the network to synchronize token dispatches across compute units.
This design faithfully models the interplay between expert routing and parallel execution in MoE serving.

\niparagraph{Expert offloading.}
While expert parallelism improves scalability, placing all experts directly on compute units is often infeasible when memory is limited, motivating a line of research on expert offloading\cite{pre-gated-moe, duplex}.
Some approaches pre-fetch experts to overlap computation with data transfer\cite{pre-gated-moe}, while others offload experts to memory-intensive compute units (e.g., PIM) to improve performance~\cite{duplex}.
%
%
To enable such studies, as shown in Fig.~\ref{fig:overview}(c), \simname implements expert offloading schemes and, to the best of our knowledge, is the first system-level simulator to provide such capability.
Researchers can flexibly configure where and how experts are offloaded, enabling exploration of diverse offloading strategies.

\input{table/system-config}

\subsection{Memory-aware Serving with Prefix Caching}
Prefix caching exploits the high reuse of input sequences in LLM inference by reusing cached prefixes, thereby significantly reducing Time to First Token (TTFT).
Recent LLM serving frameworks~\cite{vllm, zheng2024sglangefficientexecutionstructured} already support prefix caching, and active research continues to explore new caching mechanisms and policies~\cite{cacheblend}.
Despite its importance, none of the prior works models it systemically, leaving the effects of prefix caching unexamined in simulation.
This gap motivates a simulator that can capture prefix caching and its system-level implications, enabling systematic evaluation of cache policies.

\niparagraph{Prefix cache manager.} To address this gap, \simname implements prefix caching and, to the best of our knowledge, is the first system-level simulator to support this feature.
We build this feature on RadixAttention~\cite{zheng2024sglangefficientexecutionstructured} and extend the memory model with a radix tree–based prefix cache controlled by a \textit{prefix cache manager} (Fig.~\ref{fig:overview}(d)).
%
%
%
Each request triggers a longest-prefix match from radix tree, and on a prefix hit, corresponding memory-transfer events are inserted into execution trace to model loading of the blocks. 
After prefill phase, new prefixes are inserted into radix tree, and capacity pressure triggers eviction. 
In the current implementation, the radix tree uses the compute unit’s local memory (e.g., GPU or NPU memory) as the first-tier cache, with evicted blocks spilling into host CPU memory.
%
%
The hierarchy is easily reconfigured (e.g., to include SSD tiers), and \simname also supports both per-instance and global shared caches to study different caching scopes.
This flexible modeling of prefix caching allows researchers to explore diverse cache management strategies and understand their impact. 

%% file: table/system-config.tex
\begin{table}[t]
\centering
\caption{Serving configurations used for evaluation.}
\begin{tabular}{lll}
\toprule
Config & Description & Instance / GPU per Inst. \\
\midrule
S(D/M)    & Single-instance Dense/MoE & 1 inst., 1 $\times$ RTX3090 \\
M(D/M)    & Multi-instance Dense/MoE & 2 inst., 1 $\times$ RTX3090 \\
PD(D/M)    & P/D-disaggregated Dense/MoE & 2 inst., 1 $\times$ RTX3090 \\
* + PC  &  * + Prefix Cache & -- \\
\bottomrule
\end{tabular}
\label{tab:config}
\vspace{0.1em}
\end{table}

%% file: body/evaluation.tex
\section{Evaluation}

\subsection{Methodology}
\niparagraph{System baseline.}
We use two baseline systems for evaluating \simname: (1) a real GPU system equipped with four NVIDIA RTX~3090 GPUs and an Intel Xeon Gold~6326 CPU, and (2) a Google Colab system with a TPU-v6e-1 instance.
As the serving software, we employ vLLM~\cite{vllm}, a state-of-the-art framework for LLM inference. 
In our evaluation, we use two representative LLMs: Llama3.1-8B as a dense model and Phi-mini-MoE as a MoE model.
For the workload, we sample 100 requests from ShareGPT~\cite{sharegpt} and synthesize arrival patterns using a Poisson process with a rate of 10 requests per second.

\niparagraph{\simname configuration.}
To evaluate \simname, we integrate GPU and TPU backends by extracting performance models through \textit{operator-level profiler}. 
On the GPU side, we profile both LLM models on the GPU system mentioned above, configuring the simulator with matching device specifications of 24GB memory capacity, 936GB/s memory bandwidth, and interconnect modeled as PCIe 4.0$\times$16.
To further demonstrate the ease of hardware integration, we modify the profiler to run on a TPU-v6e-1 instance in Colab, configuring the simulator with an equivalent setup of 32GB of memory capacity, 1.6TB/s memory bandwidth, and an interconnect bandwidth of 800GB/s.
Table~\ref{tab:config} summarizes the serving configurations used to evaluate \simname.

\niparagraph{Simulator baseline.}
We also compare the hardware integration complexity and simulation time of \simname with previous LLMServingSim~\cite{llmservingsim}.
Specifically, we use two variants: (1) LLMServingSim with the hardware simulator enabled, and (2) LLMServingSim+, which omits hardware simulation by replaying pre-simulated results.

\subsection{Hardware Integration}

\input{table/hw-integration}
Table~\ref{tab:profiler_eval} compares the hardware integration cost of \simname against the previous LLMServingSim\cite{llmservingsim} when integrating a TPU backend, 
measured in terms of Lines of Code (LoC) excluding comment and blank lines, offline profiling time, online simulation time, and error rate compared to the real hardware.
In the case of LLMServingSim, enabling the hardware simulator requires substantial code modifications, including rewriting the compilation and simulation flow to match its requirements and even creating new APIs for usability. 
%
%
In contrast, \simname reduces the implementation effort to only 258 lines of code, mainly by just replacing CUDA APIs with XLA APIs for TPU compatibility. Compared to the 4.8K lines required by LLMServingSim, this dramatically lowers the engineering burden while offline profiling takes only 21 hours using our \textit{operator-level profiler.}
Moreover, as \simname is trace-driven, simulation time is reduced by 509$\times$, and since the profiler itself performs validation, the error rate decreases from 14.7\% to 2.25\%.
This lightweight integration flow allows researchers to seamlessly integrate novel hardware and explore a wide range of systems.

\subsection{Simulator Validation}

\begin{figure}[t]
  \centering
  \includegraphics[width=1\columnwidth]{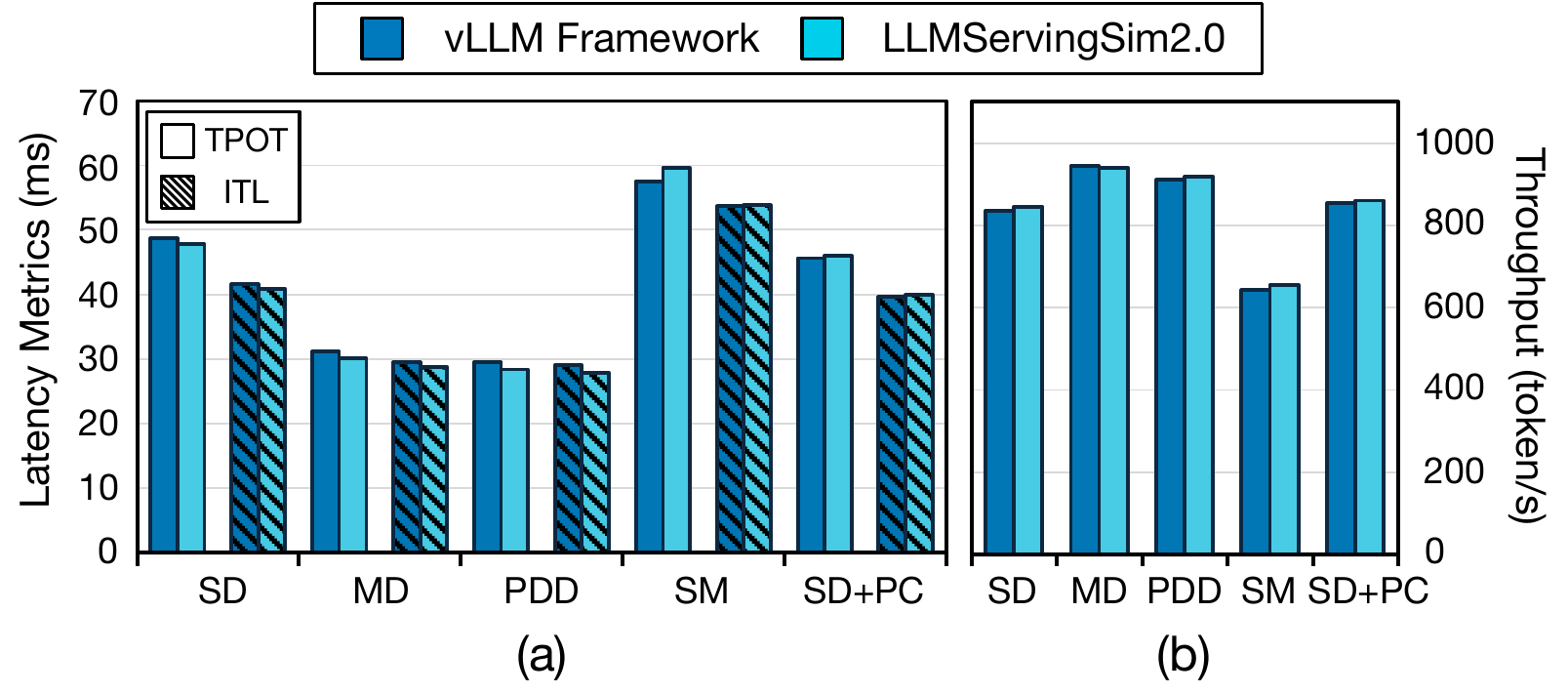}
  \vspace{-1.5em}
  \caption{Latency and throughput comparison of vLLM and \simname, across five system configurations.}
  \label{fig:validation}
\end{figure}

Fig.~\ref{fig:validation} compares \simname with vLLM across five configurations. 
Fig.~\ref{fig:validation}(a) shows average Time per Output Token (TPOT) and Inter-Token Latency (ITL), while Fig.~\ref{fig:validation}(b) shows average token generation throughput. 
Across all configurations, the simulated latency and throughput closely follow the trends observed in real GPU-based LLM serving, while the error rate remains within 5\%.
Single-instance settings show lower error rates than multi-instance settings because multi-instance deployments introduce additional variability, including request routing and network contention, which are harder to capture in a simulator. 
Within the multi-instance case, PDD incurs higher error than MD because the two instances must coordinate and exchange intermediate results. 
For MoE, the gating function routes tokens differently across layers and batches, increasing variance and making accurate reproduction more difficult. 
Despite these challenges, \simname captures key sources of variability and closely reproduces real GPU-based serving behavior, indicating that it provides a reliable abstraction of modern LLM serving systems.

\subsection{Simulation Time}

\begin{figure}[t]
  \centering
  \includegraphics[width=0.95\columnwidth]{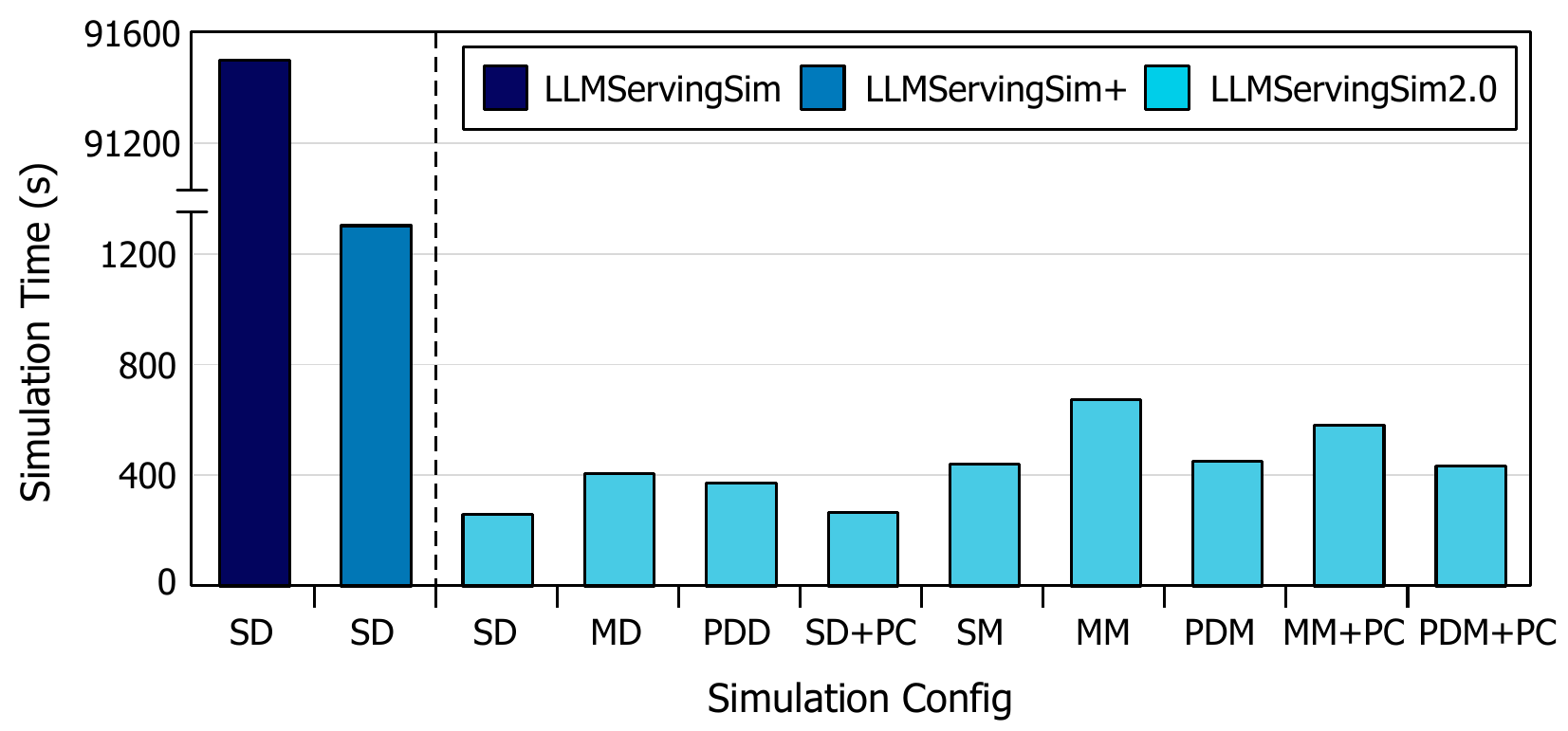}
  \vspace{-0.4em}
  \caption{Simulation time comparison of \simname against LLMServingSim and LLMServingSim+ across multiple system configurations.}
  \label{fig:simulation_time}
\end{figure}

Fig.~\ref{fig:simulation_time} reports the wall-clock time to simulate 100 ShareGPT requests. 
We run LLMServingSim2.0 across nine configurations and compare against two baselines: LLMServingSim~\cite{llmservingsim} and LLMServingSim+.
LLMServingSim incurs the longest runtime due to its hardware simulation, while LLMServingSim+ finishes much faster by removing the hardware simulation. 
For \simname, even in the slowest case (MM), \simname finishes 1.94$\times$ faster than LLMServingSim+, due to a streamlined simulation workflow that minimizes unnecessary computation.
Among the configurations, single-instance setups are fastest, followed by P/D disaggregation, and then multi-instance, with runtime increasing as system complexity grows.
MoE is slower than the dense model because each layer must perform expert routing, which introduces additional overhead.
Prefix caching, on the other hand, can either reduce or increase simulation time, since it accelerates request serving and decreases simulation iterations, but also incurs extra overhead from cache management.
Overall, simulating 100 requests in under 12 minutes shows that \simname achieves practical simulation speed, even with the additional functionality integrated into the framework.

%


%% file: table/hw-integration.tex
\begin{table}[t]
\centering
\caption{Hardware integration performance of \simname}
\begin{tabular}{lcccc}
\toprule
Simulator & LoC & Prof. Time & Sim. Time & Error \\
\midrule
LLMServingSim~\cite{llmservingsim} & 4764 & --  & 1524.7 min & 14.7$\%^{\dagger}$ \\
\simname & \textbf{258}  &  21 hr  & \textbf{3.0 min}      & \textbf{2.25\%} \\
\bottomrule
\end{tabular}
\label{tab:profiler_eval}
\vspace{0.5em}
{\footnotesize
\raggedright
$^{\dagger}$Error rate of LLMServingSim is taken from the original paper~\cite{llmservingsim}.
}
\vspace{-0.2em}
\end{table}

%% file: body/conclusion.tex
\section{Conclusion}

%
This paper introduces \simname, which brings modern LLM serving into a unified, accurate, and configurable simulator so that systems can be evaluated as they would run on scale-out deployments. 
By enabling controlled experiments on diverse hardware, disaggregation, MoE, parallelism, and caching within one framework, \simname lowers the barrier to rigorous methodology and accelerates progress in this field.
Looking ahead, we will extend support for emerging devices such as Compute Express Link (CXL) to broaden the range of research questions it can address.